# Preliminary Report: Enhancing Role Differentiation in Conversational HCI Through Chromostereopsis

**Matteo Grella**

Independent Researcher

*matteo.grella@protonmail.com*

## Abstract

We propose leveraging chromostereopsis, a perceptual phenomenon inducing depth perception through color contrast, as a novel approach to visually differentiating conversational roles in text-based AI interfaces. This method aims to implicitly communicate role hierarchy and add a subtle sense of physical space.

## Executive Summary

We introduce a novel method for enhancing conversational role differentiation and user agency in text-based conversational AI interfaces by exploiting the perceptual phenomenon of chromostereopsis, also known as the "advance–retreat effect".

Positioned at the intersection of perception science and HCI design, this approach aims to enrich user experience by implicitly communicating spatial relationships.

Building upon traditional methods that utilize color-coding and textual formatting, our approach employs a dark terminal background to enhance the optical illusion wherein warm colors, particularly red, appear to advance while cool colors, specifically blue, perceptually recede. User-submitted messages are vividly rendered in red to foreground their presence, whereas AI-generated responses utilize bright blue to suggest background positioning. Additionally, active user inputs are distinctly displayed in stark white to reinforce user agency and interaction immediacy.

Although theoretically grounded in perceptual and cognitive principles, empirical validation through controlled user studies remains necessary to assess effectiveness and visual comfort.

```
Good evening, Matteo.
> Why is my text red and yours blue?
It's chromostereopsis: red appears closer, blue farther, visually distinguishing our roles.
> Feels like my words pop forward!
Exactly—it adds a subtle sense of physical space, making our interaction feel more natural.
> Nice touch, like we're facing each other.
```

Figure 1: Example of Chromostereopsis-based Conversational Interface (HTML/CSS). Color Required for Depth Perception.

# 1. Introduction

Text-based conversational interfaces, from classic terminals to modern chatbots, traditionally employ various visual methods to differentiate between conversational roles, yet these solutions typically don't leverage explicit spatial depth cues.

While graphical user interfaces (GUIs) utilize techniques like shadows, layering, and segmenting, purely text-based systems typically rely on minimal styling options such as color coding and textual formatting for distinguishing conversational roles [Zhang & Walji, 2007]. These traditional methods primarily address clarity but do not explore perceptual depth illusions as functional design elements.

Our novel solution applies chromostereopsis, a visual illusion where specific color combinations induce an apparent spatial depth, effectively simulating three-dimensionality [Healey & Enns, 2012; Thompson et al., 1993]. By assigning distinct chromatic treatments to user messages and AI-generated texts, our method aims to implicitly communicate spatial relationships, enhancing the user's perceptual experience of conversational clarity and structure. We further emphasize the user's active engagement through a highly luminant visual treatment during active input, which perceptually situates user agency at the interface forefront.

Critically, given that this visual phenomenon is underutilized in practical interface design due to concerns over visual comfort [Thompson et al., 1993], empirical studies will be essential to assess user acceptability and effectiveness comprehensively.

# 2. Related Work

Visual depth cues, including shadows, gradients, and layering, are extensively utilized in graphical user interface design to improve readability, clarity, and interaction flow [Ware, 2012; Norman, 2013].

However, text-based interfaces traditionally utilize fewer visual dimensions, primarily restricted to typography and limited color-coding to distinguish conversational roles. Usability research on chat interfaces confirms that users respond positively to color-coded message roles, appreciating visual differentiation between participants' contributions. Yet, these methods generally treat color categorically rather than exploiting perceptual illusions to imply spatial depth.

Chromostereopsis, the perceptual phenomenon where warm colors (e.g., red) advance while cool colors (e.g., blue) recede, is documented extensively in visual perception literature [Thompson et al., 1993; Kaya & Epps, 2004; Healey & Enns, 2012]. Despite its potential for hierarchical encoding, chromostereopsis is commonly described as either visually engaging or potentially uncomfortable, with HCI research cautioning against prolonged exposure due to visual strain [Thompson et al., 1993].

Prior studies of conversational interfaces primarily emphasize readability, navigability, and user agency, typically employing categorical color-coding without depth implications. The novel aspect of our research lies precisely in repurposing chromostereopsis, a known perceptual illusion historically regarded with caution in design, as a deliberate tool for enhancing conversational role differentiation in text-based AI interactions.

## 3. Methodology

### 3.1 Psychological Foundations

Our approach relies on established psychological insights:

**Chromostereoptic Illusions**: Utilizing warm colors, notably reds, creates the illusion of advancement, while cooler colors, primarily blues, enhance the illusion of depth and recession [Kaya & Epps, 2004].

**Perceptual Hierarchy**: Leveraging perceptual phenomena aligns with cognitive principles suggesting that visual differentiation can implicitly guide user attention and enhance hierarchical comprehension [Ware, 2012].

### 3.2 Proposed Visual Schema

Implemented specifically in terminal-based conversational interfaces, our schema includes:

**User Messages**: Rendered in vivid red, capitalizing on the advancing nature of warm colors, perceptually foregrounding the user's conversational role.

**AI Messages**: Rendered in bright blue, employing chromostereopsis to imply perceptual recession, creating an intuitive conversational hierarchy.

**Active User Input**: Displayed prominently in stark white, reinforcing the user's immediate engagement and conversational control without relying significantly on luminance differences, thus avoiding potential readability issues and visual strain.

## 4. Discussion

### 4.1 Novelty

While color-coding roles in text interfaces is common, the deliberate and systematic use of perceptual illusions to communicate depth is unprecedented in conversational UI design literature: leveraging chromostereopsis effectively encodes conversational hierarchy and differentiates roles clearly, potentially improving the perceptual understanding of interactions.

Also, highlighting active inputs prominently aligns visual representation with cognitive priorities, reinforcing the user's control and immediacy within conversations.

## 4.2 Adaptive Interfaces

Given the variability in individual perceptual sensitivity to color-induced depth illusions like chromostereopsis, adaptive interface strategies should be considered. These interfaces would allow users to dynamically adjust visual parameters—such as color intensity, saturation, luminance, or hue shifts—to personalize their experience and reduce visual discomfort or fatigue.

User-customizable color profiles could enable adjustments within boundaries that preserve the chromostereoptic effects. While maintaining the fundamental warm-cool contrast necessary for depth illusion, users could select from a predefined spectrum of reds and blues, modify background colors, adjust saturation levels, or fine-tune luminance contrasts according to their comfort and preferences.

To prevent potential visual fatigue from extended exposure to chromostereopsis, implementing controls that allow users to disable or adjust the effect's intensity based on usage duration would be beneficial.

## 4.3 Limitations and Considerations

Chromostereopsis may cause perceptual strain or discomfort in prolonged use or for certain users, necessitating empirical studies to identify thresholds of usability and variable capabilities of terminal systems regarding color rendering may limit the consistency and effectiveness of depth illusions, suggesting a need for adaptive or fallback strategies.

Systematic user studies remain essential for confirming whether the depth illusion meaningfully improves user experience, structural clarity, and conversational comprehensibility without adverse effects.

## 5. Implementation

## 5.1 Color Schema Implementation

We designed our chromostereopsis-based approach using HTML/CSS, leveraging CSS variables to facilitate consistent application across interfaces while enabling straightforward adjustments based on experiments and users feedback.

Below are examples of our proposed color schemas:

```
body {
  /* Dark background gradient to enhance contrast and reduce eye strain */
  --bg-start: #0E1A2B;
  --bg-end:   #050F15;

  /* Human messages: vivid red with a subtle blue glow for enhanced effect */
  --user-color: #FF4D4D;
  --user-glow: rgba(0, 32, 160, 0.70);

  /* Machine messages: bright blue with a subtle red glow */
  --ai-color: #0050FF;
  --ai-glow: rgba(255, 0, 0, 0.10);

  /* Active input: stark white with white glow for maximum prominence */
  --prompt-color: #ffffff;
  --prompt-glow: rgba(255, 255, 255, 0.7);
}
```

This color schema represents an initial implementation based on the author's personal experience and visual perception that can be adapted for different contexts, user preferences, and accessibility requirements. The addition of contrasting glows (blue glow on red text, red glow on blue text) extends the core chromostereopsis concept by creating color boundaries that may enhance the depth illusion while potentially improving readability.

## 5.2 Accessibility and Color Vision Deficiencies

Our research considers the challenges faced by users with color vision deficiencies. Traditional chromostereopsis relies on red-blue contrast, which may be imperceptible to users with protanopia (inability to perceive red light), deuteranopia (inability to perceive green light), or tritanopia (inability to perceive blue light).

We propose specialized hypotheses of schemas -awaiting validation- that maintain perceptual depth effects while remaining accessible.

**Schema for Protanopia (Red Blindness)**

```
body {
  --bg-start: #0A121A;
  --bg-end: #050A10;

  /* Use yellow (appears bright) instead of red */
  --user-color: #FFDD00;
  --user-glow: rgba(0, 64, 160, 0.40);

  /* Use a darker blue with higher saturation */
  --ai-color: #0070DD;
  --ai-glow: rgba(255, 255, 100, 0.10);

  --prompt-color: #ffffff;
  --prompt-glow: rgba(255, 255, 255, 0.7);
}
```

This schema relies on the fact that yellow is perceived as highly luminant and "advancing" even for those with protanopia, while the darker blue will still appear to recede. The luminance contrast preserves the depth effect to some degree.

**Schema for Deuteranopia (Green Blindness)**

```
body {
  --bg-start: #0A121A;
  --bg-end: #050A10;

  /* Orange has higher perceptual brightness */
  --user-color: #FF9500;
  --user-glow: rgba(0, 64, 160, 0.40);

  /* Navy blue with high saturation */
  --ai-color: #0066CC;
  --ai-glow: rgba(255, 200, 0, 0.15);

  --prompt-color: #ffffff;
  --prompt-glow: rgba(255, 255, 255, 0.7);
}
```

**Schema for Tritanopia (Blue Blindness)**

```
body {
  --bg-start: #121212;
  --bg-end: #080808;

  /* Bright red still works */
  --user-color: #FF3333;
  --user-glow: rgba(0, 0, 0, 0.40);

  /* Use darker yellow instead of blue */
  --ai-color: #CCAA00;
  --ai-glow: rgba(100, 0, 0, 0.15);

  --prompt-color: #ffffff;
  --prompt-glow: rgba(255, 255, 255, 0.7);
}
```

**Universal Design Approach**

For environments where user color vision status is unknown or mixed, we recommend a universal design approach that relies primarily on luminance and saturation differences:

```
body {
  --bg-start: #0A0A0F;
  --bg-end: #050505;

  /* High luminance, high saturation */
  --user-color: #FFAA00; /* Orange-yellow appears bright to most users */
  --user-glow: rgba(0, 0, 0, 0.50); /* Dark glow for contrast */

  /* Lower luminance, high saturation */
```

```
  --ai-color: #0088CC; /* Cyan-blue ap-
pears to recede for most users */
  --ai-glow: rgba(255, 255, 255, 0.15);
/* Light glow increases visibility */

  --prompt-color: #ffffff;
  --prompt-glow: rgba(255, 255, 255,
0.7);
}
```

This universal design relies on three principles:

**Luminance Contrast**: Higher luminance elements (user messages) will appear to advance while lower luminance elements (AI messages) will appear to recede, regardless of color perception.

**Saturation Differentiation**: Highly saturated colors create stronger visual weight and tend to draw attention forward.

**Complementary Glows**: Using opposite brightness glows (dark glow on bright text, light glow on dark text) enhances visibility while maintaining the depth effect.

### 5.3 Alternative Schema Variations

For users without color vision deficiencies, we provide additional schemas examples for different use contexts:

**High Contrast Schema**

```
body {
  --bg-start: #000000;
  --bg-end: #121212;

  --user-color: #FF3333;
  --user-glow: rgba(0, 0, 255, 0.60);

  --ai-color: #3399FF;
  --ai-glow: rgba(255, 0, 0, 0.15);

  --prompt-color: #ffffff;
  --prompt-glow: rgba(255, 255, 255,
0.8);
}
```

**Subtle Schema (Reduced Eye Strain)**

```
body {
  --bg-start: #1A2632;
  --bg-end: #0A1520;

  --user-color: #FF6666;
  --user-glow: rgba(0, 32, 160, 0.40);

  --ai-color: #5580FF;
  --ai-glow: rgba(255, 0, 0, 0.07);

  --prompt-color: #f0f0f0;
  --prompt-glow: rgba(255, 255, 255,
0.5);
}
```

### 5.4 Implementation Recommendations

When implementing chromostereopsis in conversational interfaces, several practical considerations can enhance the user experience while mitigating potential drawbacks. We suggest the following approaches:

**Adaptive Configuration**: Design interfaces that detect and respond to user needs. This means checking system accessibility settings and automatically selecting appropriate color schemas – for instance, switching to the protanopia-friendly schema when color blindness settings are detected.

**Personalization Options**: Beyond predefined schemas, users should be able to fine-tune their experience. Simple sliders for adjusting color intensity or contrast can significantly impact comfort, particularly during extended use. These preferences should persist between sessions.

**Evidence-Based Refinement**: Collect qualitative and quantitative feedback through targeted user testing with diverse participants. Special attention should be paid to those with color vision deficiencies, as their experiences will reveal limi-tations in the current approach and inspire more inclusive alternatives.

**Environmental Awareness**: Consider how changing viewing conditions affect the perception of chromostereopsis. Bright sunlight on a screen requires different color intensity than a dark room. Interfaces could leverage ambient light sensors to make subtle adjustments that maintain the depth effect across environments.

**Fatigue Prevention**: Extended exposure to chromostereopsis may cause visual strain. Implementing gradual reduction in color intensity over time (similar to night mode features) can help maintain comfort during longer sessions without disrupting the user's workflow.

## 6. Conclusion

We have proposed a novel perceptual approach to improve role differentiation in text-based conversational interfaces by leveraging the optical illusion of chromostereopsis.

By strategically assigning role-specific colors based on established perceptual depth cues, our approach creates visual differentiation that implies spatial hierarchy and emphasizes user agency through subtle perceptual effects rather than explicit interface elements.

The color schemas we have presented—from standard implementations to accessible alternatives—provide a comprehensive framework for exploring this concept across diverse user populations and interface contexts.

While our approach is theoretically grounded in established perceptual and cognitive principles, its practical application to conversational interfaces represents uncharted territory in HCI research: we actively encourage researchers and designers to implement, test, and refine these schemas, collecting empirical data that can further our understanding of how perceptual illusions might enhance conversational interface design.

Through collaborative exploration and iterative improvement, this preliminary work may lead to more intuitive, accessible, and perceptually rich conversational experiences - or may reveal that traditional approaches without perceptual depth effects ultimately serve users better.